# Towards an open geotechnical data platform in France
## Vers une plateforme de données géotechniques ouvertes en France

I. Halfon*, M. Beaufils, *BRGM, F-45060 Orléans, France*

*i.halfon@brgm.fr

**ABSTRACT:** An important quantity of geotechnical data is constantly collected for all the new projects of civil engineering. This data includes generally core and destructive boreholes, in which samples are taken for laboratory testing and in situ geotechnical tests are performed. The density of geotechnical data is particularly high in urban areas. The data is collected by geotechnical engineering or drilling Companies for public or private owners. Data is essential to define the geotechnical conditions of a project and are obviously necessary for the geotechnical design of foundations or underground structures. However, most of the time, data is not accessible in a numerical format, and at the end of the project, is often forgotten, while it could be reused for neighbouring projects. It is a great loss of information and knowledge to the technical and scientific geosciences community, and it represents a significant cost for the country economy. In France, the BRGM (French geological survey office) is currently developing a new open access platform dedicated to the capitalization and accessibility of geotechnical data, compliant with the FAIR principles (Findable, Accessible, Interoperable and Reusable). This paper describes the challenges to overcome and the possible solutions, considering the high diversity of geotechnical tests, explains the need of keeping the exhaustive data set for each test, and describes the next stages of development.

**RÉSUMÉ**: Une quantité importante de données géotechniques est constamment collectée pour tous les nouveaux projets de génie civil. Ces données comprennent généralement des sondages carottés et destructifs, dans lesquels des échantillons sont prélevés pour des essais en laboratoire et des essais géotechniques in situ sont réalisés. La densité des données géotechniques est particulièrement élevée dans les zones urbaines. Les données sont collectées par des sociétés d'ingénierie géotechnique ou de forage pour le compte de maitres d'ouvrage publics ou privés. Les données sont essentielles pour définir les conditions géotechniques d'un projet et sont évidemment nécessaires pour la conception géotechnique des fondations ou des infrastructures. Mais la plupart du temps, les données ne sont pas accessibles sous format numérique, et à la fin du projet, elles sont souvent oubliées, alors qu'elles pourraient être réutilisées pour des projets voisins. C'est une grande perte d'informations pour la communauté technique et scientifique des géosciences, et cela représente un coût important pour l'économie du pays. En France, le BRGM (Bureau de Recherches Géologiques et Minières) développe actuellement une nouvelle plateforme de science ouverte dédiée à la capitalisation et à l'accessibilité des données géotechniques, conforme aux principes FAIR (Findable, Accessible, Interoperable and Reusable). Cet article décrit les difficultés techniques et solutions proposées, compte tenu de la grande diversité des essais géotechniques, explique la nécessité de conserver l'ensemble des données brutes pour chaque essai, détaille les différentes étapes du développement.

## 1 INTRODUCTION

Geotechnical engineering encompasses the study of the mechanical properties of soils and rocks and their interaction with the structure or development to be built. The vast majority of geotechnical studies require the acquisition of a certain amount of data, including in particular the stratigraphic succession, the lithological description of the soils, the measurement of their identification parameters, the measurement of their mechanical behaviour parameters (resistance, deformability, consolidation, etc.). The density of geotechnical data is particularly high in urban areas.

The data includes generally core and destructive boreholes, with eventually samples for laboratory testing and in situ geotechnical tests.

For each project, one or more drilling contractors, laboratories and geotechnical engineers, acquire this data in one or more investigation stages as part of an order placed by the project owner.

The data is then analysed and interpreted by a geotechnical engineer. They produce a report, which presents the geotechnical model of the project derived from the investigations carried out, the geotechnical parameters interpreted and the various justifications for the geotechnical structure under investigation. The raw data is generally provided as an appendix to this report, in the form of borehole logs, photographs of samples or cores, graphs, test reports, results tables, etc. This report and its appendices are attached to the tender documents for the works companies, who can also supplement them and make their own interpretations.

Once the work has been completed, the geotechnical reports and the data they contain are archived by the various parties involved. More often than not, if the structure does not behave abnormally, this data is never reused, or even forgotten altogether.

The advent of digital tools (databases, GIS, etc.) now makes it possible to capitalise on this data, and opens up the possibility of re-using it for neighbouring projects, for example, or quite simply to improve knowledge of the subsoil, facilitate and rationalise its development (Kokkala and Marinos, 2022), (Pando et al., 2022).

In France, the BRGM (French geological survey office) is currently developing a new open access platform dedicated to the capitalization and accessibility of geotechnical data, compliant with the FAIR principles (Findable, Accessible, Interoperable and Reusable).

## 2 ABOUT GEOTECHNICAL DATA

### 2.1 Nature, heterogeneity, provenance and reusability

Geotechnical data is both numerous and diverse in nature. It may be textual data (e.g. lithological descriptions or geotechnical classification symbols), single physical quantities (e.g. water content or density), pairs or groups of linked physical quantities (e.g. shear strength, which is made up of cohesion and friction angle), or temporal data (e.g. piezometric records).

It is also important to note the existence in geotechnics, as in other scientific disciplines, of a rich and sometimes ambiguous vocabulary to designate the different properties of soils and rocks, with terms of close physical meanings. For example, we speak of undrained cohesion, or undrained shear strength, or sometimes apparent cohesion, or even short-term cohesion. It is therefore essential to define a precise lexical reference system as part of a geotechnical data bank.

In addition, some geotechnical parameters are measured directly (e.g. the cone resistance of the static penetrometer). Others (most of them in fact) are the result of an in situ or laboratory test, involving raw measurements, parameters linked to the test or to the measuring device, intermediate calculations and interpretation by the engineer, leading in the end to interpreted parameters. For instance, in the Ménard pressuremeter test, the contextual data are the depth, the groundwater level, the probe characteristics and the calibration parameters. The raw data are the probe volume measurements at each step of pressure, and at different times. The intermediate parameters are the corrected pressure-volume measurements and the $p_1$ and $p_2$ pressures of the beginning and the end of the pseudo-elastic phase. And the final interpreted data are the net limit pressure (Pl*), the net creep pressure (Pf*) and the pressuremeter modulus ($E_M$).

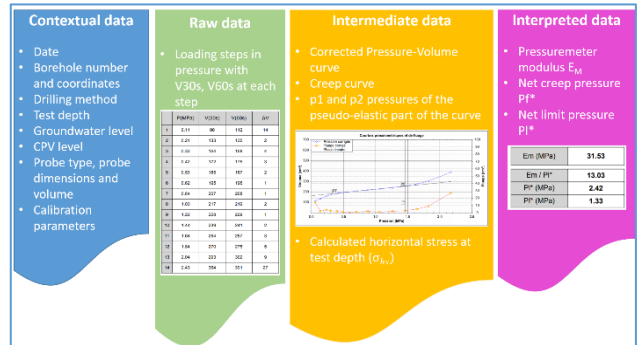

*Figure 1. Different classes of data, example of the Ménard pressuremeter test (©BRGM).*

The capitalisation of geotechnical parameters in a database therefore raises the question of what data do we want to store? One option would be to store only the final, interpreted data from a geotechnical test (e.g. Pl*, Pf* and $E_M$ in the case of the previous example). However, this would appear to be a very restrictive choice: in fact, these parameters are of little use to the geotechnical engineer if he does not have all the contextual elements of the test (drilling method and diameter, probe dimensions and characteristics, volume-pressure raw data and curves, creep curve, etc.).

It is therefore essential to know and control the source of geotechnical data and, to do this, to be able to store in the database all the data from a geotechnical test: contextual data, raw measurements, curves, intermediate calculations and, of course, the interpreted values. This choice of exhaustiveness gives the engineer who re-uses the data the opportunity to assess the quality of the test and to make his own interpretation. On the other hand, this choice implies a number of constraints and requirements: the management of a large quantity of data, the use of a harmonised vocabulary for all geotechnical data, and the need of using structured and standardized data, i.e. data that obeys precise classification rules, and finally.

### 2.2 Data standardization and interoperability

Data structuration is a prerequisite for setting up a database. This essential step requires the use of an unambiguous vocabulary for the various physical properties that are stored. To ensure that this vocabulary is understood and shared by all geotechnical engineers, the logical choice is to draw on geotechnical standards (testing standards and Eurocode 7). Each term used for a physical quantity must be accompanied by a clear definition. Any synonyms must be listed and associated with this property.

In order to facilitate data description and to harmonise semantic, BRGM has established registers and acting as a reference point for geoscience vocabularies (https://data.geoscience.fr/ncl/). The registers published cover a wide range of earth sciences topics, including lithology, chronostratigraphy, fossils, minerals, as well as methods, observed properties and units of measurement. They include many entries relating to Geotechnics (in-situ tests, laboratory tests, geomechanical properties and associated units). Each register includes additional information such as synonyms, semantic relationships (hierarchies, equivalences, genealogies, etc.), as well as other useful information for subject specialists such as symmetrical relationships and replacements for obsolete terms.

For Geotechnics, the BRGM registry provides a list of ObservedProperties, Procedure. They have been drawn up as part of an open and collaborative science approach, so that any user can propose modifications or additions to the proposed definitions.

Interoperability is "the ability of a product or system, whose interfaces are fully known, to work with other existing or future products or systems without restrictions on access or implementation" (Wikipedia). Interoperability is also essential if data is to be easily exchanged, either to feed the database or to export data. It requires the development of data standards for all kind of geotechnical data, which is still not the case at present time.

### 2.3 Digital transition: BIM, Digital Twin, Geomodeling

Since 2018, BRGM has been looking to the issue of standardising geotechnical data. The opportunity to tackle this topic was offered by the national project MINnD, which main aim was to extend the capacities of OpenBIM standards, in particular the Industry Foundation Class (IFC) format from bSI, for describing infrastructures in their environment (Beaufils & al, 2020), (Rives & al, 2020).

For geotechnical engineering, the work was carried out in collaboration with several professional bodies in France, including ANDRA, CAN, CETU, EDF, EGIS, GEOLITHE, GEOS, SETEC, SYSTRA, TERRASOL and VINCI CONCESSIONS. There was a consensus to build the extension for geotechnics on existing standards (e.g. NF P 94-500) or existing recommendations (e.g. AFTES GT32).

Having as a target the definition of a data exchange format with common semantics between GIS and BIM tools, the group focused on identifying the data exchanged, describing it and organising it.

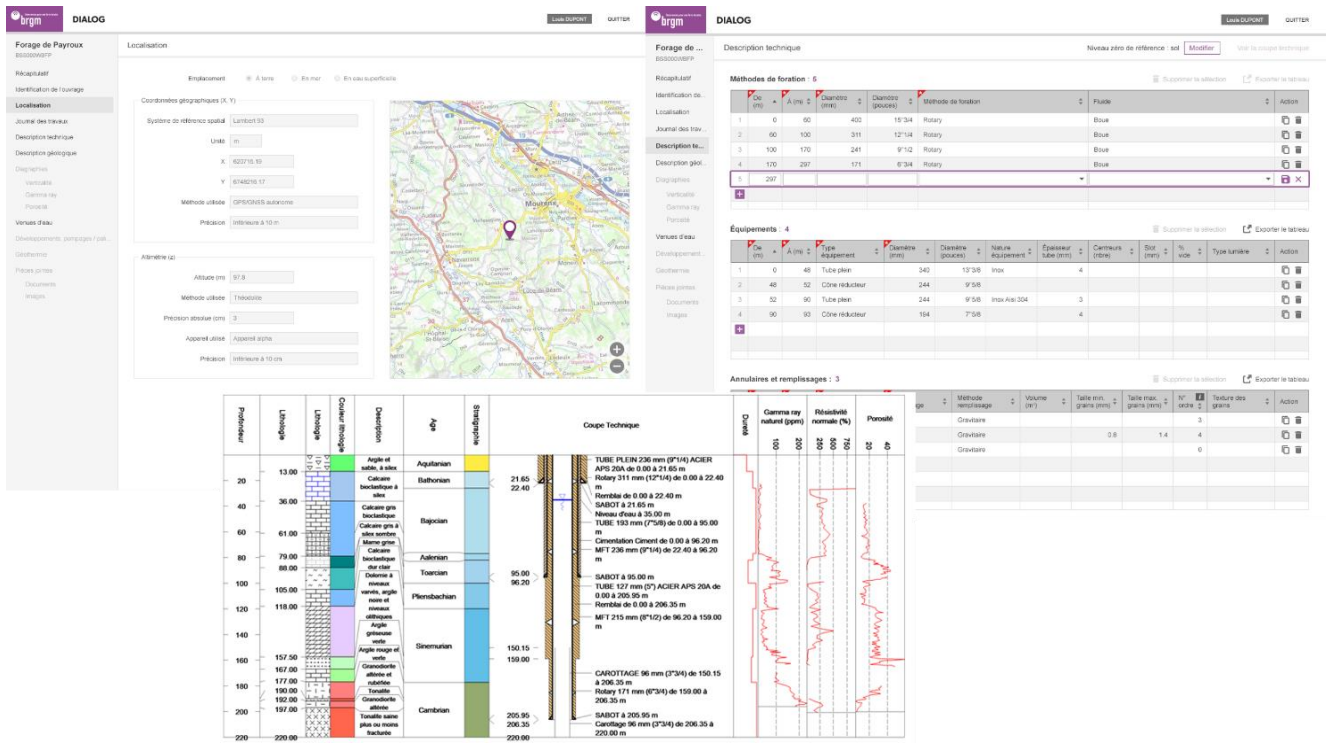

*Figure 2: DIALOG web application screenshots (©BRGM).*

The openGIS standards promoted by the Open Geospatial Consortium (OGC), used in particular for geology and hydrogeology, served as the basis for this work. Also the group was aware of the existence of the AGS/AGSi (Chadwick & al, 2019) and DIGGS (Bachus & al, 2020) standards and in February 2022 a project called the Geotech Interoperability Experiment was launched (Beaufils & al, 2022).

Again lead by BRGM, the project proposed to federate the geotech community around standards for BIM and GIS.

## 3 STATUS OF GEOTECHNICAL DATA MANAGEMENT IN FRANCE

### 3.1 Rules and laws

In France, geotechnical data is the property of the organisation that commissions and pays for it. French law, set out in the Mining Code, requires the declaration to the Authorities of any drilling deeper than 10 m, whatever its purpose (geological, hydrogeological, geotechnical, geothermal investigation, water pumping, etc.).

This declaration is made using a web application (DUPLOS) developed by BRGM. It must be made before the drilling is carried out. An identification number is then created by BRGM and sent to the organisation that made the declaration. The Mining Code also stipulates that all documents relating to the drilling can then be sent to BRGM. To this end, DIALOG, a web application has been developed to enable the lithological and stratigraphic description of the borehole, previously declared, to be entered. To date, this application cannot receive geotechnical properties, except a pdf file attachment.

### 3.2 BRGM Survey about a national geotechnical data base

Before starting to develop an open access national platform for geotechnical data, BRGM launched a survey to the French geotechnical community. The objectives of this survey were to: find out about the practices of French geotechnical engineers with regard to archiving and exchanging geotechnical data, assess the opinion of geotechnical engineers for data sharing, collect the expectations and needs of future users of the platform, and also their criticisms or warnings.

The survey took place in mid-2023. It was relayed by various geotechnical associations and more than 200 answers were collected A slight majority (53%) are representatives of geotechnical engineering companies, 20% are civil works or foundation contractors, 14% are engineers from public organisations, 6% are public owners. This high level of participation demonstrates the interest of French geotechnical community in this subject.

It appears that a large majority of respondents (85%) consult regularly or occasionally the BSS (subsoil database, which at date is a borehole database, with litho-stratigraphic data only).

With regard to archiving practices, digital archiving of geotechnical reports and data is widely used (92% for reports and 81% for geotechnical data). In the case of data, pdf format is used in the majority (53%). Re-usable formats (such as text, excel or csv) are used by 31% of respondents, and more structured formats (such as AGS, BIM or others) accounting for only 13%.

Support for the project from the geotechnical community is relatively good, as shown by the scores between 1 and 5 given to different proposals, presented in Figure 3. There is also a strong demand for all the raw values in order to be able to make full use of this database.

| About a national openaccess geotechnical database, on a score from 1 (disagree) to 5 (agree), you would say …. | Score |
|---|---|
| This would make it easy to retrieve old data | 4,6 / 5 |
| This would make it possible to avoid carrying out (or repeating) investigations in areas where data is already available | 3,6 / 5 |
| This would help to better define new or future geotechnical investigations | 4,6 / 5 |
| This would be useful, provided that the raw data from the trials is accessible. | 4,2 / 5 |
| This is of limited use, as geotechnical investigation is always necessary anyway. | 2,7 / 5 |
| The benefits are limited, as we have no control over the quality or reliability of the tests carried out by others. | 3,0 / 5 |
| The benefits are limited, as it takes a lot of time to archive or consult data. | 2,1 / 5 |
| This is likely to reduce the market for geotechnical surveys. | 2,5 / 5 |

*Figure 3. Results of the BRGM survey about geotechnical database.*

However, some concerns were expressed. For example the question of the reliability/quality of the data made available on the platform, and the risks of reducing the market for geotechnical surveys in France.

## 4 NATIONAL GEOTECHNICAL DATA PLATFORM – TECHNICAL ASPECTS

### 4.1 User profiles

Users of the platform will of course be geotechnical professionals (engineers, civil works contractors), but also infrastructure project owners, teachers, students or simply individuals looking for information about the subsoil.

The data suppliers will be the public or private project owners who order for geotechnical investigations. They may delegate the supply of data files to their project managers or geotechnical sub-contractors.

## 4.2 Structured and unstructured data

The geotechnical data that will be collected can be classified into two categories: structured data and unstructured data.

Structured data is data that has already been organised in a digital format, which may be a proprietary format (for example the files used for acquisition equipment), a format specific to geotechnics (such as AGS or DIGGS), a format associated to the further application (ifc for BIM, or RESQML for geological modelling applications) or a csv or excel format with predefined fields. In order to be able to collect and store this structured data, it is necessary to make a mapping that provides the link between a given parameter in the database and the same parameter in the original file.

The unstructured data are the geotechnical test sheets only available in paper or pdf format. For these data, several possibilities will be provided: downloading the scanned document or pdf file, or manual entry of the test. Further developments based on AI are explored to extract key data in a pdf file.

## 4.3 Functionalities to develop

Data input interfaces will be developed, on the principle of the DIALOG application currently used to enter lithostratigraphic descriptions of boreholes. Ultimately, the goal is to welcome the largest possible variety of geotechnical tests. However, the tables for each type of test will be developed progressively so as to quickly have an operational tool for the most common tests (e.g. pressuremeter tests, which are very common in France), which will gradually be enhanced with the other types of test. As mentioned above, the content of these tables should include all the fields corresponding to contextual, raw, intermediate and interpreted data.

For users, the data will be accessible via a cartographic interface, with borehole points displayed, as in situ or laboratory tests are always attached to an investigation location. Data will be able to be exported in a variety of formats. Queries by geographical sector, by type of test or by geological unit will be developed.

Finally, connections with geotechnical tools and software are also envisaged: for example, with 2D or 3D geomodelling tools.

## 4.4 OpenGeotech: the project and its architecture

A pilot project, called OpenGeotech, is currently under development at BRGM. Aimed at giving substance to the standardisation work undertaken previously, it proposes to:

- extend the DIALOG application to the collection and description of geotechnical tests,
- have those data made available through an a standardized protocol called OGC SensorThings API, as recommended and promoted in the Geotech Interoperability Experiment.

In addition to illustrating and validating the standard activity previously mentionned, these two combined actions are done in order to facilitate the process of geotechnical data collection and also offer a „reward" or „money back" to the declarant by offering him the capacity to acess and reuse the declared data in other softwares (geomodeling, soil structure calculus or BIM tool) thanks to the interoperable API.

At the current stage of development, drilling parameters (instantaneous logs) and three types of test have been targeted: the Menard Pressuremeter test, the Standard Penetration Test (SPT) and the Cone Penetration Test (CPT).

The description of more tests is planned to be offered in the future, including Laboratory Tests and Samples / Specimen description. These actions will be carried on as part of a larger project in BRGM, currently known as the Geotech Information System project.

## 5 CONCLUSION AND PERSPECTIVES

Until now, in France, geotechnical data acquired during investigations is not usually reused. Although digital archiving is increasingly used by geotechnical engineers, the data is very rarely organised in a structured way.

The creation of a structured database enabling geotechnical data to be banked and shared with the public is a task that has been entrusted to BRGM by the authorities. It aims to enhance the value of the data acquired, to reduce the need to carry out new surveys unnecessarily, to better target the investigations to be carried out and to improve knowledge of the French subsoil.

A survey of French geotechnical professionals was launched in mid-2023. Most of them were in favour of the project, although they did express some concerns about the quality, reliability and confidentiality of the data.

From a technical point of view, the wealth and complexity of geotechnical data raises a number of questions about the choice of data to be banked, as well as the methods for feeding and consuming the data, and the exchange formats. BRGM has chosen to provide for the possibility of storing all the raw test data, so that future users can assess the quality of the data and re-analyse the tests.

A demonstrator called OpenGeotech is currently being developed in-house at BRGM. A first public version will be available by the end of 2024.